\documentclass[onecolumn,showpacs,showkeys,amsmath,amssymb]{revtex4}
\usepackage{times}

\newcommand{\beq}{\begin{equation}}
\newcommand{\eeq}{\end{equation}}

\newcommand{\ba}{\begin{eqnarray}}
\newcommand{\ea}{\end{eqnarray}}

\def\t{\theta}

\def\ve{\varepsilon}
\def\vt{\vartheta}

\def\gs{\mathrel{\lower0.6ex\hbox{$\buildrel {\textstyle >}\over{\scriptstyle \sim}$}}}
\def\ls{\mathrel{\lower0.6ex\hbox{$\buildrel {\textstyle <}\over{\scriptstyle \sim}$}}}

\begin{document}

\title{Analytical Kerr black hole lensing in the weak deflection limit}

\author{Mauro Sereno$^a$}
\email{sereno@physik.unizh.ch}

\author{Fabiana De Luca$^{a,b}$}
\email{fadeluca@physik.unizh.ch}

\affiliation{
$^a$
Institut f\"{u}r Theoretische Physik, Universit\"{a}t Z\"{u}rich,
Winterthurerstrasse 190, CH-8057 Z\"{u}rich, Switzerland
$^b$ Dipartimento di Fisica `E.R. Caianiello', Universit\`{a} di Salerno, via Allende, I-84081 Baronissi (SA), Italy.
}

\date{October 4, 2006}

\begin{abstract}
We present an analytical treatment of gravitational lensing by a Kerr black hole in the weak deflection limit. Lightlike geodesics are expanded as a Taylor series up to and including third-order terms in $m/b$ and $a/b$, where $m$ is the black hole mass, $a$ the angular momentum and $b$ the impact parameter of the light ray. Positions and magnifications of individual images are computed with a perturbative analysis. At this order, the degeneracy with the translated Schwarzschild lens is broken. The critical curve is still a circle displaced from the black hole position in the equatorial direction and the corresponding caustic is point-like. The degeneracy between the black hole spin and its inclination relative to the observer is broken through the angular coordinates of the perturbed images.
\end{abstract}


\maketitle

\section{Introduction}

The gravitational deviation suffered by photons passing near massive compact bodies provided one of the first observational tests of general relativity and is still considered as an excellent probe for gravity theories. Black hole lensing has been emerging as a pretty promising tool for gravitational investigations in both weak and strong fields. On the observational side, interest in this topic is mainly motivated by the super-massive black hole supposed to be hosted in the radio source Sgr A* in the Galactic center. Planned high-resolution observations at the astrometric resolution of the microarcosecond ($\mu$as) should allow in the next future a clean detection of higher order effects in gravitational lensing. It is now well understood that a photon passing near a  black hole can suffer either a strong or a weak deflection. The latter occurs when the minimum distance is much larger than the gravitational radius. The former occurs when photons wind around the black hole making one or more loops and producing images very near to the shadow. 

Analytical treatments have been worked out for generic spherically symmetric spacetimes, just assuming that the light ray follows the geodesic equations. The deflection angle always diverges logarithmically when the minimum impact parameter is in the very neighborhood of complete capture \cite{boz02}, whereas for larger distances and in the weak deflection limit a Taylor series expansion works pretty well \cite{ke+pe05}. Some investigations interpolating between the two limits have been also performed \cite{am+ar06}.

Whereas lensing by either Schwarzschild or generic spherically symmetric black holes has been extensively investigated, a full analytical description of lensing by a Kerr Black hole is still missing. An intrinsic angular momentum breaks the spherical symmetry heavily affecting the gravitational field. The modern era in the study of Kerr geodesics came when Carter \cite{car68} was able to fully separate the Hamilton-Jacobi equation. Based on this technique of separation of variables, many later investigations addressed the light propagation near a rotating body \cite{cha83} and numerical studies flourished. \citet{cu+ba73} considered the optical appearance of a point source in a circular orbit in the equatorial plane of an extreme Kerr black hole. \citet{vie93} investigated the image positions through a code based on the quasi-analytic solution of the geodesic problem by elliptical integrals. \citet{ra+bl94} provided a detailed analysis of the optical structure of the primary caustic surface. The analytical extension of the strong deflection limit methodology to the Kerr black hole has been performed as well in the limit of small values of the angular momentum and for sources in the asymptotically flat region of the space-time \cite{boz03,boz+al05,boz+al06}. A full description of caustics and the inversion of lens mapping for sources near them has been performed up to the second order in $a$. 

Several analytical investigations in the weak deflection limit, considering the first correction due to the angular momentum, were performed in the past. The null tetrad formalism of geometrical optics was used to study the optical properties of images into the field of an axisymmetric system \cite{pi+ro77}. \citet{ep+sh80} performed a calculation based on the post-Newtonian expansion. The motion equations for two spinning point-like particles, when the spin and the mass of one of the particles were zero, were resolved in  \cite{ib+ma82,iba83} by expanding the Kerr metric in a power series of the gravitational constant $G$. \citet{bra86} evaluated the equations of motion for a light ray in the weak deflection limit up to and including second order corrections in $m/b$ e $a/b$, where $m$ is the black hole mass, $a$ the angular momentum and $b$ is the impact parameter. \citet{dym86} discussed the additional time-delay due to rotation by integrating the light geodesics. \citet{gli99} considered light rays passing outside a spinning star in the framework of  the Lense-Thirring metric. Kopeikin and collaborators \citep{kop97,ko+sc99,ko+ma02} investigated the gravito-magnetic effects in the propagation of light in the field of self-gravitating spinning bodies. \citet{as+ka00} considered the light deflection angle caused by an extended, slowly rotating lens. These analyses where then extended to generic spinning mass distributions, in the usual framework of gravitational
lensing theory \citep{ser02,se+ca02,ser03,ser03b,cap+al03,ser05}, i.e. {\it i)}
weak field and slow motion approximation for the lens and {\it ii)}
thin lens hypothesis. Expressions for bending and time delay of electromagnetic waves were found for stationary rotating deflectors with general mass distributions.

At second order in $m/b$ and $a/b$, the Kerr lens is observationally equivalent to the Schwarzschild one because of the invariance under the global translation of the center of the lens mass \cite{as+ka00}. In this paper we take a step forward and study the lensing up to the next order. Following \citet{bra86}, we start from the lightlike null geodesics and then move to the gravitational lensing for a configuration in which both source and observer lie in the asymptotically flat region of the spacetime. We take care of expressing the results in terms of the invariants of the light ray, avoiding ambiguities connected to coordinate-dependent quantities \cite{ke+pe05,bo+wi03}. 

The paper is organized as follows. In Section~\ref{sec:basi}, we introduce our notation and recall some properties of the Kerr spacetime. In Section~\ref{sec:lens}, the lens equations in the weak deflection limit are derived starting from the geodesics. Section \ref{sec:imag} is devoted to the solution of the lens equations with a perturbative method. Sections~\ref{sec:magn} and \ref{sec:crit} discuss the magnification of the images and the singularity of the lens mapping, respectively. Section~\ref{sec:conc} is devoted to some considerations. Finally, Appendixes \ref{app:a} and \ref{app:b} report some details on the evaluation of the integrals appearing in the geodesic equations. In this paper, we will use units $G=c=1$, with $c$ the light speed in the vacuum.

\section{Basics}
\label{sec:basi}

The Kerr black hole metric in the Boyer-Lindquist coordinates, $\{ t, r, \vartheta, \phi\}$, is given by
\beq
\label{ker1}
ds^2= \left( 1- \frac{r_\mathrm{Sch} r}{\rho^2}\right)dt^2-\frac{\rho^2}{\Delta}dr^2 -\rho^2 d \vartheta^2-
\left( r^2 +a^2 +\frac{a^2 r_\mathrm{Sch} r \sin^2 \vartheta }{\rho^2}\right)\sin^2 \vartheta d\phi^2 
+\frac{a r_\mathrm{Sch} r \sin^2 \vartheta}{\rho^2}dt d\phi ,
\eeq
where
\begin{eqnarray}
\rho^2 & \equiv & r^2 +a^2  \cos^2 \vartheta , \label{ker2} \\
\Delta & \equiv & r^2 + r_\mathrm{Sch} r+a^2 .\label{ker3}
\end{eqnarray}
The constant $r_\mathrm{Sch} = 2 m$ is the gravitational radius.  We consider a static observer and a static emitter in the asymptotically flat region of the spacetime, $r \gg r_\mathrm{Sch}$. The observer coordinates are denoted $\{r_\mathrm{o}, \vartheta_\mathrm{o}, \phi_\mathrm{o} =0 \}$, where $\phi_\mathrm{o}$ has been fixed without loss of generality. The source coordinates are denoted as $\{r_\mathrm{s}, \vartheta_\mathrm{s}, \phi_\mathrm{s} \}$. In what follows, we will also use the modified polar coordinate $\mu \equiv \cos \vt$. The null geodesics for a light ray can be expressed in terms of the first integrals of motion $J$ and $Q$ \cite{car68,cha83}. The photon trajectory from the source to the observer can then be written as
\begin{eqnarray}
\pm \int \frac{d r}{\sqrt{R}} & = & \pm \int \frac{ d \vartheta}{\sqrt{\Theta}} \label{ker4} , \\
-\phi_\mathrm{s} & = &\int \frac{a \left(r_\text{Sch}r-a J\right )}{\pm \Delta{\sqrt{R}} }dr +  \int \frac{ J d \vartheta}{\pm \sin^2 \vartheta \sqrt{\Theta} } \label{ker5} ,
\end{eqnarray}
where
\begin{eqnarray}
R(r) &  \equiv & r^4+\left(a^2-J^2-Q \right) r^2+\left[ (J-a)^2+Q\right] r_\text{Sch} r-a^2 Q , \label{ker6} \\
\Theta (\vartheta) & \equiv & \left(a^2-J^2 \csc ^2 \vartheta \right) \cos ^2\vartheta +Q . \label{ker7}
\end{eqnarray}
The signs of $\sqrt{R}$ and $\sqrt{\Theta}$ are adhered to the signs of $d r$ and $d\vartheta$, respectively. The signs change at the inversion points in the $r$ or $\vartheta$-motion. We consider the standard framework of gravitational lensing, where the source of radiation and the observer are remote from the lens. In such a configuration the equatorial plane is crossed at least once, so that the range $Q<0$ is excluded in our analysis. 

Along his path from the source to the observer, the photon passes by the black hole at a minimum distance $r_\mathrm{min}$ which, in the weak deflection limit, is much larger than the gravitational radius. This distance of closest approach is the only turning point in the $r$-motion. Differently from the strong deflection limit, when the photon may perform several oscillations around the equatorial plane, in the weak deflection limit there is just one inversion point also in the angular polar motion. $\vartheta$ can attain either a maximum or a minimum depending on the direction taken by the photon starting from $\vt_\mathrm{s}$. If $\vt$ is initially growing ($\mu$ decreasing), the polar motion will attain a maximum $\vt_\mathrm{max}$ (a minimum $\mu$) and then it will decrease to get to the observer at $\vt_\mathrm{o}$ ($\mu$ increases to get to $\mu_\mathrm{o}$), otherwise for an initially decreasing $\vt$.

The light ray minimum radial distance $r_\mathrm{min}$ to the lens is determined by $R(r)=0$, whose roots represent inversion points in the radial motion. In general, there can be up to four real roots with $r_\mathrm{min}$ being the largest one. In the weak deflection limit, there is just one inversion point at a distance of order of the impact parameter \cite{bra86}. The impact parameter is an invariant of motion defined geometrically by the perpendicular distance from the center of the lens to the asymptotic tangent line to the light ray at the observer. For the spherically symmetric case it reduces to $\sqrt{J^2+Q}$. A fundamental assumption in the weak deflection limit is that the point of closest approach lies well outside the gravitational radius, i.e. $r_\mathrm{Sch} \ll \sqrt{J^2+Q}$.
Let us now introduce two independent expansion parameters in terms of the invariants of motion
\begin{eqnarray}
\epsilon_{m} & = & \frac{m}{\sqrt{J^2+Q}} , \label{esp1} \\
\epsilon_{a} & = & \frac{a}{\sqrt{J^2+Q}} . \label{esp2}
\end{eqnarray}
In what follows, we will expand quantities of interest in both $\epsilon_{m}$ and $\epsilon_{a}$. For the sake of brevity, we will refer to terms of order ${\cal{O}}(\epsilon_{m}^i \epsilon_{a}^j)$ as terms of order ${\cal{O}}(\epsilon^n)$ with $n = i +j$ . We will produce our results up to a given formal order in $\epsilon$, collecting terms of a given order in $\epsilon$ coming from any combination of the two quantities $\epsilon_a$ and $\epsilon_m$. We recall how terms like ${\cal{O}} (a ~m)$ or ${\cal{O}} (m^2)$, which according to our notation are both of formal order $\sim \epsilon^2$, are not necessarily of the same physical order. This is the case only for a maximal (or nearly maximal) Kerr black hole, when $|a| \sim m $.

Let us find the minimum radial distance in the weak deflection limit. We can solve the equation $R (r_\mathrm{min}) =0$ expressing $r_\mathrm{min}$ as a power series in $\epsilon$. We then find
\begin{eqnarray}
\label{inv2}
r_\mathrm{min} & \simeq & (J^2+Q)^{1/2} \left\{
1 - \frac{r_\text{Sch}}{2 \sqrt{J^2+Q}}  -\frac{a^2 J^2}{2\left(J^2+Q\right)^2}   +   \frac{a r_{\text{Sch}} J}{\left(J^2+Q\right)^{3/2}}  -    \frac{3 r_\text{Sch}^2}{8 \left(J^2+Q\right)} \right. \\
& - &  \frac{r_{\text{Sch}}^3}{2 \left(J^2+Q\right)^{3/2}}     +    \frac{3 a J r_{\text{Sch}}^2}{2 \left(J^2+Q\right)^2}  -    \frac{J^2 a^2 r_{\text{Sch}}}{\left(J^2+Q\right)^{5/2}} \nonumber \\
& - & \left. \frac{ J^2 (J^2-4  Q) a^4}{8 \left(J^2+Q\right)^4}   +   \frac{J \left(J^2-Q\right) r_{\text{Sch}} a^3}{\left(J^2+Q\right)^{7/2}}   +    \frac{\left(8 Q-51 J^2\right) r_\text{Sch}^2 a^2}{16 \left( J^2+Q \right)^3}+\frac{3 J r_\text{Sch}^3 a}{ \left( J^2+Q \right)^{5/2}}  -   \frac{105 r_\text{Sch}^4}{128 \left( J^2+Q \right)^2}   +{\cal{O}}(\epsilon^5)
 \right\}  .\nonumber
\end{eqnarray}
An expression for the minimum approach including terms ${\cal{O}}(\epsilon^3)$ can be found in \cite{bra86}. Eq.~(\ref{inv2}) for the spherical symmetric case ($a=0$) agrees with the result in \cite{ke+pe05}. The observer and the source lie very far from the lens in the asymptotically flat region of the spacetime. It can be shown that $b/r_\mathrm{o} \sim b/r_\mathrm{s} \sim \epsilon_m$ \cite{ke+pe05}. This scaling relation will be useful when collecting terms $r_\mathrm{min}/r_\mathrm{o}$ and $r_\mathrm{min}/r_\mathrm{s}$ in the integrals. In what follows, without loss of generality, we will consider non negative values of the spin $a$.

\section{Lens equations}
\label{sec:lens}

The geodesic equations, Eqs.~(\ref{ker4},~\ref{ker5}) will provide the lens equations. They can be viewed as a map between the angular position of the source, $\{ \mu_\mathrm{s}, \phi_\mathrm{s}\}$, and the image position, which is a function of the couple of invariants $\{J, Q \} $. Details on the resolution of the radial and angular integrals in the weak deflection limit are given in Appendix~\ref{app:a} and \ref{app:b}, respectively. Following \cite{bra86}, we can recast the geodesic equation in a quite compact form. The first equation, Eq.~(\ref{ker4}), provides a description of the polar motion. It can be written as
\beq
\label{lens1}
\mu_\mathrm{s} = -\mu_\mathrm{o} \cos \delta +(-1)^k \sin \delta \left( \frac{Q}{J^2+Q}-\mu_\mathrm{o}^2\right)^{1/2} ,
\eeq
where
\begin{eqnarray}
\label{lens2}
\delta & = & \frac{2 r_{\text{Sch}}}{\sqrt{J^2+Q}} + \frac{15 \pi  r_{\text{Sch}}^2}{16 \left(J^2+Q\right)}-\frac{4 J r_{\text{Sch}}a}{\left(J^2+Q\right)^{3/2}} \\
& + & \frac{16 r_{\text{Sch}}^3}{3 \left(J^2+Q\right)^{3/2}}-\frac{15 \pi J a r_{\text{Sch}}^2}{4 \left(J^2+Q\right)^2}
+\left( \mu_\mathrm{o}^2+ \frac{5 J^2-3 Q}{J^2+Q} +\frac{Q}{\frac{J^2 \mu_\mathrm{o}^2}{\mu_\mathrm{o}^2-1}+Q} \right) \frac{ r_{\text{Sch}}a^2}{(J^2+Q)^{3/2}}
\nonumber \\
&-& (J^2+Q)^{1/2} \frac{r_\mathrm{o}+r_\mathrm{s}}{r_\mathrm{o} r_\mathrm{s}}  -\frac{(J^2+Q)^{3/2}}{6} \frac{r_\mathrm{o}^3+r_\mathrm{s}^3}{r_\mathrm{o}^3 r_\mathrm{s}^3} +\frac{\sqrt{J^2+Q} \mu _\mathrm{o}^2 \left(1-\mu _\mathrm{o}^2\right)}{2 \left(J^2 \mu _\mathrm{o}^2-Q \left(1-\mu _\mathrm{o}^2\right)\right)}\frac{r_\mathrm{o}+r_\mathrm{s}}{r_\mathrm{o} r_\mathrm{s}}a^2 
+{\cal{O}}(\epsilon^4) .
\nonumber
\end{eqnarray}
Up to terms $\sim \epsilon^2$, $\delta$ was already evaluated in \cite{bra86}. For a null angular momentum ($a=0$) and for very distant source and observer, $r_\mathrm{o}, r_\mathrm{s} \rightarrow \infty$, the parameter $\delta$ reduces to the deflection angle induced by the Schwarzschild black hole \cite{ke+pe05}. The parameter $k$ in Eq.~(\ref{lens1}) accounts for the direction in the polar motion of the photon at the observer. $k$ is even (odd)  if $\vt$ attains $\vt_\mathrm{max}$ ($\vt_\mathrm{min}$), i.e for photons coming from below (above) the black hole.

The second geodesic equation,  Eq.~(\ref{ker5}), accounts for the azimuthal motion. Using Eq.~(\ref{lens1}), we can write
\beq
\label{lens3}
-\phi_\mathrm{s}= \frac{J }{|J|}\pi +
\frac{J \delta}{\sqrt{J^2+Q}}\frac{1}{1-\mu _\mathrm{o}^2} 
\left[ 1 - (-1)^k \delta  \frac{\mu _\mathrm{o}}{\sqrt{1-\mu _\mathrm{o}^2}}\sqrt{\frac{Q}{J^2+Q}-\mu_\mathrm{o}^2} \right]
  +\frac{2 a r_{\text{Sch}}}{J^2+Q}+\delta \phi_\mathrm{s} ,
\eeq
where $\delta \phi_\mathrm{s}$ is a contribution of order $\sim \epsilon^3$ ,
\ba
\label{lens4}
\delta \phi_\mathrm{s} & \equiv  & \frac{5 \pi  r_{\text{Sch}}^2 a}{4 \left(J^2+Q\right)^{3/2}}-\frac{8 J r_{\text{Sch}}^3 \left(2 \left(J^2+Q\right) \mu _\mathrm{o}^4+\left(2 J^2-Q\right) \mu_\mathrm{o}^2-Q\right)}{3 \left(J^2+Q\right)^3 \left(1-\mu _\mathrm{o}^2\right)^3} - \left(\frac{Q}{Q-\left(J^2+Q\right) \mu_\mathrm{o}^2}+3 \right) \frac{J r_{\text{Sch}}a^2}{( J^2+Q )^2} \nonumber 
\\
& - & \frac{a^2 J \mu _\mathrm{o}^2}{2 \left(J^2 \mu _\mathrm{o}^2-Q \left(1-\mu _\mathrm{o}^2\right)\right)} \frac{r_\mathrm{o}+r_\mathrm{s}}{r_\mathrm{o} r_\mathrm{s}} +
\frac{J \left(2 J^2 \mu _\mathrm{o}^2 \left(\mu_\mathrm{o}^2+1\right)-Q \left(-2 \mu _\mathrm{o}^4+\mu _\mathrm{o}^2+1\right)\right)}{\left(J^2+Q\right) \left(1-\mu _\mathrm{o}^2\right)^3}
\left[
\frac{1}{3} \left(J^2+Q\right) \frac{\left(r_\mathrm{o}+r_\mathrm{s}\right)^2}{r_\mathrm{o}^2 r_\mathrm{s}^2} \right. \nonumber \\
& -& \left. \frac{2 \left(r_\mathrm{o}+r_\mathrm{s}\right) r_{\text{Sch}}}{r_\mathrm{o} r_\mathrm{s}} + \frac{4
   r_{\text{Sch}}^2}{J^2+Q} \right] \frac{r_\mathrm{o}+r_\mathrm{s}}{r_\mathrm{o} r_\mathrm{s}}.
\ea
Up to including terms of order of ${\cal{O}}(\epsilon^2)$, Eqs.~(\ref{lens3},~\ref{lens4}) have been already evaluated in \cite{bra86}.

The lens equation are usually given in terms of the apparent angular position of the image onto the plane of the sky (POS), i.e. the coordinate angles $\{ \theta_1, \theta_2\}$, and of the angular position of the source in absence of the lens, $\{ B_1, B_2\}$. In the asymptotic flat region, the Boyer-Lindquist coordinates can be thought as spherical coordinates. We can introduce a Cartesian observer coordinate system centered on the black hole, with the  $x_3$-axis running along the line of sight (LOS), i.e. the line from the observer to the lens, and the $x_2$-axis tracing the projection of the spin axis onto the POS. Then, $\t_1$ and $\t_2$ are measured along the $x_1$- and the $x_2$-axis, respectively. In other words, $r_\mathrm{o} \tan  \theta_1$  and $r_\mathrm{o} \tan  \theta_2$ are the apparent (length-)displacement of the image perpendicular to the projected axis of symmetry of the black hole and the apparent (length-)displacement perpendicular to the equatorial plane in the sense of the angular momentum of the black hole, respectively. The angles $\{ \theta_1, \theta_2\}$ are strictly linked to the invariants of motion and to the impact parameter through the relations
\begin{eqnarray}
r_\text{o}\frac{\tan \t_1 }{\sqrt{1+\tan^2 \t} }& = & -\frac{J}{\sqrt{1-\mu_\mathrm{o}^2}} \label{lens5} \\
r_\text{o}\frac{\tan \t_2 }{\sqrt{1+\tan^2 \t} }& = & -(-1)^k \sqrt{Q + a^2 \mu_\mathrm{o}^2 - J^2\frac{\mu_\mathrm{o}^2}{1-\mu_\mathrm{o}^2}},  \label{lens6}
\end{eqnarray}
with $\t$ being the angular separation of the image from the black hole, $\tan^2 \theta = \tan^2 \theta_1 +\tan^2 \theta_2$. The parameter $k$ can be expressed in terms of $\t_2$ as
\beq
(-1)^k = -\frac{\t_2}{|\t_2|}.
\eeq
Equations~(\ref{lens5},~\ref{lens6}) can be obtained by defining the tangent to the ray at the observer through the equations of motion of the photon. Photons are named prograde (retrograde) if they turn on the equatorial plane in the same (opposite) sense of the black hole. Prograde photons ($J>0, Q=0$) are seen by the observer on the left side of the black hole ($\t_1 <0$). The relation between the angular position of the image and the impact parameter for a spherically symmetric lens, $b=\sqrt{J^2+Q} = r_\mathrm{o} \sin \t$, can be easily recovered from Eqs.~(\ref{lens5},~\ref{lens6})

The angular position of the source  $\{ B_1, B_2\}$  can be expressed in terms of the Boyer-Lindquist coordinates. Considering the intercept of the line through the source and the observer with the POS, we find
\ba
D_\mathrm{s} \tan B_1 & = &\sin \phi _\mathrm{s}  r_\mathrm{s} \sqrt{1-\mu _\mathrm{s}^2} \label{lens7} ,\\
D_\mathrm{s} \tan B_2 & = & r_\mathrm{s} \left( \mu_\mathrm{s} \sqrt{1-\mu _\mathrm{o}^2} -  \mu _\mathrm{o} \sqrt{1-\mu _\mathrm{s}^2} \cos \phi _\mathrm{s} \right)  \label{lens8},
\ea
where $D_\mathrm{s}$ is the distance along the LOS from the observer to the plane of the source, i.e. the plane through the source and perpendicular to the LOS. The relations between the radial coordinates and the distances measured along the LOS are
\ba
D_\mathrm{d} & = & r_\mathrm{o}, \\
D_\mathrm{ds} & = & - r_\mathrm{s} \left( \mu _\mathrm{o} \mu _\mathrm{s} + \cos \phi _\mathrm{s}\sqrt{(1-\mu _\mathrm{o}^2 ) (1-\mu _\mathrm{s}^2)}  \right) , \label{lens9} \\
D_\mathrm{s} & = & D_\mathrm{d} + D_\mathrm{ds} .
\ea
$D_\mathrm{ds}$ is the distance along the LOS between the lens plane and the source plane. The $D_i$ distances must be properly intended as angular diameter distances. The relations in Eqs.~(\ref{lens5},~\ref{lens6}) and Eqs.~(\ref{lens7},~\ref{lens8}) allows us to put the geodesics, Eqs.~(\ref{lens1},~\ref{lens3}), in the classical form of the lens mapping
\ba
B_1 & =  & B_1(\t_1,\t_2), \label{lens10}  \\
B_2 & =  & B_2(\t_1,\t_2) .\label{lens11} 
\ea
Once we use angular coordinates for the image positions instead on the invariants of motion, it can be appropriate to introduce a series expansion parameter in the weak deflection limit based on the angular Einstein ring, 
\beq
\label{lens19}
\theta_\mathrm{E} \equiv \sqrt{2 r_\mathrm{Sch} \frac{r_\mathrm{s}}{r_\mathrm{o}(r_\mathrm{o}+r_\mathrm{s})}} .
\eeq
Following \cite{ke+pe05}, we define
\beq
\label{lens20}
\varepsilon \equiv  \frac{\theta_\mathrm{E}}{4 D}   ,
\eeq
where $D \equiv r_\mathrm{s}/(r_\mathrm{o}+r_\mathrm{s})$. We remark as differently from usual analyses in the weak deflection limit, we are adopting radial distances in the definition of the Einstein radius instead of the distances measured along the LOS. Differences are of order of $\varepsilon^3$. Angles can be rescaled in terms of $\theta_\mathrm{E}$. We then assume that the solution of the lens equations can be written as a series in $\ve$,
\ba
\t_1 & = & \t_\mathrm{E} \left\{ \t_{1(0)} +\t_{1(1)} \ve + \t_{1(2)} \ve^2 +{\cal{O}}(\ve^3) \right\}, \\
\t_2 & = & \t_\mathrm{E} \left\{ \t_{2(0)} +\t_{2(1)} \ve + \t_{2(2)} \ve^2 +{\cal{O}}(\ve^3) \right\}, \\
\t   & = & \t_\mathrm{E} \left\{ \t_{(0)}  +\t_{(1)}  \ve + \t_{(2)}  \ve^2 +{\cal{O}}(\ve^3)\right\} .
\ea
The above expressions must be read with the same caveats we discussed about the parameter $\epsilon$ after introducing the proper expansion parameters $\epsilon_m$ and $\epsilon_a$ in Eqs.~(\ref{esp1},~\ref{esp2}). As a matter of fact, a coefficient of the form $\t_{(i)}$ will be written as a polynomial of $i$-th order in $a/m$, collecting terms which are not necessarily of the same order.

 The source position can be rescaled defining
\beq
\beta_i \equiv \frac{\tan B_i}{\t_\mathrm{E}} .
\eeq
Including terms up to $\sim \ve^2$, the lens equations take the very simple form
\ba
B_1 & =  & \t_1 - D \hat{\alpha}_1 (\t_1,\t_2)  ,  \label{lens12}      \\
B_2 & =  & \t_2 - D \hat{\alpha}_2 (\t_1,\t_2)  ,  \label{lens13}  
\ea
where $\hat{\alpha}$ is the bending angle, defined as the angle between the asymptotic direction of the light ray at the observer and the asymptotic direction at the emitter. At order $\sim \ve^3$, equations become more complicated. The deflection angle is an invariant of motion and can be expressed in terms of the constants $J$ and $Q$ together with the mass and the spin of the black hole. On the other hand, by its own definition, it does not depend on the source and observer positions once they lie in the very asymptotic region of spacetime. The source position can then be directly related to the deflection angle considering source and observer at an infinite distance. This allows us to neglect higher order contributions to the path of the light ray near the black hole.
%, i.e. assuming that the intercept of the asymptotic incoming and outgoing light rays belongs to the lens plane. 
The two components of the deflection angle can then be evaluated considering the geodesics for source and observer Eqs.~(\ref{lens1},~\ref{lens2}) at $r_\mathrm{o}, r_\mathrm{s} \rightarrow \infty$, through the equation
\beq
\label{lens14}
\tan B_i (\mu_\mathrm{s}(r_\mathrm{o}, r_\mathrm{s} \rightarrow \infty),\phi_\mathrm{s} (r_\mathrm{o}, r_\mathrm{s} \rightarrow \infty))= - \frac{D_\mathrm{ds}}{D_\mathrm{s}} \tan \hat{\alpha}_i.
\eeq
It is simple to verify that for an equatorial motion, $\mu_\mathrm{s}=\mu_\mathrm{o}=0$, Eq.~(\ref{lens14}) reduces to the well known $\hat{\alpha} = \Delta \phi -\pi$. For the Kerr black hole, we get
\ba
\hat{\alpha}_1 & = &
 2 \frac{r_{\text{Sch}}}{b} \frac{b_1}{b} 
+  \frac{15 \pi}{16} \left( \frac{r_{\text{Sch}}}{b}\right)^2 \frac{b_1}{b} 
-\frac{2 \left(b_2^2-b_1^2\right) r_{\text{Sch}} a \sqrt{1-\mu _\mathrm{o}^2}}{b^4}
+8\left[ 1 -\frac{1}{3}\left( \frac{b_1}{b}\right)^2 \right] \left( \frac{r_{\text{Sch}}}{b}\right)^3  \frac{b_1}{b}   \label{lens15} \\
& -  & 
\left[ \frac{2 \left(b_2^2-b_1^2\right) \left(1-\mu _\mathrm{o}^2\right)}{b^2}+1\right] \frac{a^2 r_{\text{Sch}} }{b^3} \frac{b_1}{b}
-\left[ \frac{5 \pi  \sqrt{1-\mu_\mathrm{o}^2} \left(b_2^2-2 b_1^2\right)}{4 b^2} - \frac{4 b_2 \mu _\mathrm{o}}{b}\right] \frac{a r_{\text{Sch}}^2}{b^3} +{\cal{O}}(\epsilon^4) , \nonumber
\\
\hat{\alpha}_2 & = &
2\frac{r_{\text{Sch}}}{b} \frac{b_2}{b} 
+\frac{15 \pi}{16} \left( \frac{r_{\text{Sch}}}{b}\right)^2 \frac{b_2}{b} +\frac{4 b_1 b_2 a r_{\text{Sch}} \sqrt{1-\mu _\mathrm{o}^2}}{b^4}
+8 \left[ 1-\frac{b_2^2}{3 b^2}\right] \left( \frac{r_{\text{Sch}}}{b}\right)^3\frac{b_2}{b}  , 
\label{lens16}\\
& + &
\left(\frac{15 \pi b_1 b_2 \sqrt{1-\mu _\mathrm{o}^2}}{4 b^2} -\frac{4 b_1 \mu _\mathrm{o}}{b}\right ) \frac{a r_\text{Sch}^2}{b^3}
+
\left[ \left(\frac{b_2}{b} \right)^2  \left( \left(\frac{b_1}{b_2}\right)^2 -1 \right)^2  \mu _\mathrm{o}^2 -2 \frac{ b_1^2 (3-2 \mu _\mathrm{o}^2) -2 b_2^2}{b^2} \right] \frac{a^2 r_\text{Sch} }{b^3} \frac{b_2}{b}  
+{\cal{O}}(\epsilon^4)  , \nonumber
\ea
where inspired by Eqs.~(\ref{lens5},~\ref{lens6}) we have introduced the parameters
\begin{eqnarray}
b_1  & \equiv &  - \frac{J}{\sqrt{1-\mu_\mathrm{o}^2}} \label{lens17} \\
b_2  & \equiv & - (-1)^k \sqrt{Q - J^2\frac{\mu_\mathrm{o}^2}{1-\mu_\mathrm{o}^2}},  \label{lens18}
\end{eqnarray}
and $b =\sqrt{b1^2+b_2^2} =\sqrt{J^2+Q}$. For the spherically symmetric Schwarzschild black hole, Eqs.~(\ref{lens15},~\ref{lens16}) agree with the result in \cite{ke+pe05}. The spin enters in the deflection angle only if coupled with the mass. A first attempt to evaluate the term proportional to $m^2 a$ in the deflection angle was already performed in \cite{asa+al03}. We remark as in the derivation in \cite{asa+al03}, some higher order geometrical terms are neglected or, in other words, angles are identified with their tangents. This can affect the relation between the impact parameter and the distance of closest approach. The discussion of the equatorial motion, $b_2=0, \mu_\mathrm{o}=0$ is enough to understand some features of how the spin affects the deflection angle. We have
\beq
\hat{\alpha}_1 = 
2 \frac{r_{\text{Sch}}}{b} \frac{b_1}{b} 
+ \frac{15 \pi}{16} \left( \frac{r_{\text{Sch}}}{b}\right)^2 \frac{b_1}{b} 
+2 \frac{ a r_{\text{Sch}} }{b^2}
+ \frac{16}{3}  \left( \frac{r_{\text{Sch}}}{b}\right)^3  \frac{b_1}{b} 
+ \frac{a^2 r_{\text{Sch}} }{b^3} \frac{b_1}{b}
+ \frac{5 \pi}{2} \frac{a r_{\text{Sch}}^2}{b^3} +{\cal{O}}(\epsilon^4) .
\eeq
Whereas the gravito-electric field is always attractive, the gravito-magnetic field attracts towards the black hole only photons which move in the equatorial plane in the opposite sense of the spinning lens ($b_1 > 0$) . Terms directly proportional to the angular momentum $a$ are strictly related to the dragging of inertial frames and then act differentially on opposite sides of the hole. The deflection angle is enhanced for retrograde photons ($b_1 > 0$) and reduced for prograde photons ($b_1 < 0$). The term proportional to $a^2$ is instead related to the quadrupolar distortion caused by the black hole spin \cite{ra+bl94}. It just perturbs the spherical symmetry of the system but it does not act differentially. 

To give some numerical estimates, let us consider Sgr A* in the Galactic center, at nearly 8~Kpc from the Sun, which should host a supermassive black hole with mass $\sim 3.6 \times 10^6 M_\odot$ \cite{eis+al05}. The minimum distance of orbiting stars from the central black hole is $\gs 100$~AU, nearly 1500 times greater than the Schwarzschild radius, so that such sources can be considered in the asymptotic region of the spacetime. The Einstein radius corresponding to such a configuration is $\sim 0.5~m$as, i.e nearly 4~AU ($\sim 50~r_\mathrm{Sch}$) at the distance of Sgr A*. Let us consider a light ray in the equatorial plane with an impact parameter of $\sim 50~r_\mathrm{Sch}$. The total deflection angle is $\sim 4\times 10^{-2}$ radians, so that the weak deflection limit still holds. The size of the contribution to the deflection due to the dragging term $\propto a~m$ ($\propto a~m^2$) is $\sim 80(a/m)$~as ($6(a/m$)~as). The contribution of the term $\propto a^2 m$ is $\sim 0.8 (a/m)^2$~as. We see that corrections are sizeable even for low values of the angular momentum.

\section{Image positions}
\label{sec:imag}

Lens equations can be solved term by term. At the first order in the deflection angle, Kerr lensing is pure Schwarzschild lensing. The lens equations take the standard form
\ba
\beta_1  & = &\t_{1(0)}\left( 1-\frac{1 }{\theta _{(0)}^2 }\right) ,\\
\beta_2  & = &\t_{2(0)}\left( 1-\frac{1 }{\theta _{(0)}^2 }\right) ,
\ea
where $\theta _{(0)} =  \sqrt{ \theta _{1(0)}^2 + \theta _{2(0)}^2}$, with the usual solutions
\ba
\theta _{1(0)} ^{\pm} & = & \frac{1}{2} \left(1 \pm \sqrt{1 + \frac{4}{\beta ^2}} \right) \beta _1  ,\\
\theta_{2(0)} ^{\pm} & = &  \frac{1}{2} \left(1 \pm \sqrt{1 + \frac{4}{\beta^ 2}} \right) \beta_2  ,
\ea
with $\beta^2 \equiv \beta_1^2+\beta_2^2$.

%\ba
%0& =& -\frac{15 \theta _{1(0)} \left|\theta _{(0)}\right| \pi }{16}+
%2 \theta _{1(0)} \theta _{2(0)} \theta _{2(1)}+\theta _{1(1)} \left(\left|\theta
%   _{(o)}\right|^4+\left|\theta _{(o)}\right|^2-2 \theta _{2(0)}^2\right) +a_m \sqrt%{1-\mu _\mathrm{o}^2} \left(\theta _{2(0)}^2-\theta _{1(0)}^2\right)\\
%0& =&
%-\frac{15 \theta _{2(0)} \pi  \left|\theta _{(o)}\right|}{16}+2 \theta _{1(0)} \thet%a _{1(1)} \theta _{2(0)}+\theta _{2(1)}
%   \left(\left|\theta _{(o)}\right|^4-\left|\theta _{(o)}\right|^2+2 \theta _{2(0)}^2%\right)-2 a_m \theta _{1(0)} \theta _{2(0)} \sqrt{1-\mu _\mathrm{o}^2}
%\ea

The first contribution of the angular momentum appears at the next order in $\ve$. The second order terms of the solution read 
\ba
\theta _{1(1)} & = & \theta _{(1)}^\mathrm{Sch}  \frac{ \theta _{1(0)}}{\theta _{(0)}} 
+ \frac{(1-\theta _{1(0)}^2+\theta _{2(0)}^2 ) a_m \sqrt{1-\mu _\mathrm{o}^2}}{1-\theta
   _{(0)}^4} , \\
\theta _{2(1)} & = & \theta _{(1)}^\mathrm{Sch}\frac{\theta _{2(0)}}{\theta _{(0)}}-\frac{2  \theta _{1(0)} \theta _{2(0)}a_m \sqrt{1-\mu _\mathrm{o}^2}}{1-\theta
   _{(0)}^4} ,
\ea
where $a_m \equiv a/m$ and with
\beq
\theta _{(1)}^\mathrm{Sch}  =   \frac{15 \pi }{ 16 ( 1 + \t_{(0)}^2 ) } . 
\eeq
At this order there is a full degeneracy between a Kerr black hole and a Schwarzschild black hole displaced from the center along the equatorial plane in $\{ \t_1, \t_2\} \simeq \t_\mathrm{E} \{ a \sqrt{1-\mu_\mathrm{o}^2} \ve ,0 \}$. The lens equations are degenerate as well with those of a binary point-like lens with very short separation. Then, at this order, the line joining the perturbed images always goes through the `shifted' Schwarzschild lens. What happens in the POS is that, due to a positive angular momentum, the two images are apparently counterclockwisely rotated about the line of sight through the centre with respect to the line passing through the near unperturbed image produced in the Schwarzschild case \cite{ser03}.

Suppose a source at a distance $r_\mathrm{s} \sim 10$~pc beyond the supermassive black hole in the Galactic center and an Earth-based observer. Then $\t_\mathrm{E} \sim 0.07 (r_\mathrm{s}/10~\mathrm{pc})^{1/2}$~as and $\varepsilon \sim 0.76 \times 10^{-4} (r_\mathrm{s}/10~\mathrm{pc})^{-1/2}$. The shift to the image positions due to the dragging of inertial frames turns out to be of order of $\sim 4 (a/m) \mu$as, at the reach of future astrometric missions.

Equations at the third order become quite long, but solutions can be still put in a compact form. We have
\ba
\theta _{1(2)} & = &
\theta _{(2)}^\mathrm{Sch} \frac{\theta _{1(0)}}{\theta _{(0)}} +  \frac{ 16 D^2}{3} \theta _{1(0)} \theta _{2(0)}^2 \\
& + & a_m \left\{ -\frac{4 \theta _{2(0)} \mu _\mathrm{o}}{\theta _{(0)}^2} + \frac{5 \sqrt{1-\mu _\mathrm{o}^2} \pi }{16 \theta _{(0)}^3 (1-\theta _{(0)}^2 ) (1+\theta_{(0)}^2)^3}  \left[ \theta _{(0)}^2 (1+\theta _{(0)}^2 )^2 (1+4 \theta _{(0)}^2)-(12 \theta _{(0)}^6+5 \theta _{(0)}^4+4 \theta_{(0)}^2-1 ) \theta _{1(0)}^2 \right] \right\} ,\nonumber \\
& + & a_m^2 
\left\{ -\theta _{1(0)} \left(1-\mu _\mathrm{o}^2\right) \left[ \frac{4 \left(\theta _{(0)}^4+\theta _{(0)}^2+1\right) \theta _{2(0)}^2}{\left(1-\theta _{(0)}^2\right)^2 \left(\theta_{(0)}^2+1\right)^3} -\frac{\theta _{(0)}^2}{\left(\theta _{(0)}^2+1\right)^3}\right]
\right\}   , \nonumber \\
\theta _{2(2)} & = &
\theta _{(2)}^\mathrm{Sch} \frac{\theta _{2(0)}}{\theta _{(0)}} +  \frac{ 16 D^2}{3} \theta _{1(0)}^2 \theta _{2(0)} \\
& + & a_m
\left\{ \frac{4 \theta _{1(0)} \mu _\mathrm{o}}{\theta _{(0)}^2} +\frac{5 \left(-12 \theta _{(0)}^6-5 \theta _{(0)}^4-4 \theta _{(0)}^2+1\right) \theta _{1(0)} \theta _{2(0)}
   \sqrt{1-\mu _\mathrm{o}^2} \pi }{16 \theta _{(0)}^3 \left(1-\theta _{(0)}^2\right) \left(\theta _{(0)}^2+1\right)^3}\right\} 
\nonumber \\
& + & a_m^2 \left\{\frac{\theta _{2(0)} \left(\theta _{(0)}^2 \left(3 \theta _{(0)}^4+2 \theta _{(0)}^2+3\right)-4 \left(\theta _{(0)}^4+\theta _{(0)}^2+1\right) \theta_{2(0)}^2\right) \left(1-\mu _\mathrm{o}^2\right)}{\left(1-\theta _{(0)}^2\right)^2 \left(\theta _{(0)}^2+1\right)^3}\right\} , \nonumber
\ea
where
\ba
\theta _{(2)}^\mathrm{Sch} & = &
-\frac{225  \pi ^2 \left(2 \theta _{(0)}^2+1\right)}{256 \theta _{(0)} \left(\theta _{(0)}^2+1\right)^3} \\
& - & \frac{8 \left[3  \left(\theta _{(0)}^4-\theta _{(0)}^2-1\right) r_\mathrm{o}^2-3 r_\mathrm{s} \left(\theta _{(0)}^4-\theta _{(0)}^2+3\right) r_\mathrm{o}+r_\mathrm{s}^2 \left(2 \theta _{(0)}^6-7\theta _{(0)}^4+6 \theta _{(0)}^2-6\right)\right] }{3 (r_\mathrm{o}+r_\mathrm{s})^2 \theta _{(0)} (1+\theta _{(0)}^2)} .
\ea
At this order, images are no longer lined up on a line passing for a fixed position. The intercept with the axis of abscissae depends on the source position. This proves that the degeneracy between a Kerr lens and a displaced Schwratzscild lens gets lost. For a source at $r_\mathrm{s}$ beyond Sgr A*, the shift at this order to the image position due to the spin is $\sim (a/m)^i{\cal{O}}(\theta_\mathrm{E}~\varepsilon^2) \sim 0.3 (a/m)^i(10~\mathrm{pc}/r_\mathrm{s})^{1/2} p$as, with $i=1$ when considering the higher order correction due to the dragging and $i=2$ when considering quadrupolar distortion. For $r_\mathrm{s} \sim 100$~AU and $a \ls m$, we get a shift of $\sim 4\times 10^{-2} \mu$as, near the accuracy requirement for the space mission project MAXIM \footnote{http://maxim.gsfc.nasa.gov.}.

The angular distance of an image from the black hole has coefficients
\ba
\t_{(1)} & = & 
\t_{(1)}^\mathrm{Sch}+\frac{a_m \theta _{1(0)} \sqrt{1-\mu _\mathrm{o}^2}}{\theta _{(0)} \left(\theta _{(0)}^2+1\right)} , \\
\t_{(2)} & = & \t_{(2)}^\mathrm{Sch}+
\frac{5 a_m \left(4 \theta _{(0)}^4+2 \theta _{(0)}^2+1\right) \theta _{1(0)} \sqrt{1-\mu _\mathrm{o}^2} \pi }{8 \theta _{(0)}^2 \left(\theta _{(0)}^2+1\right)^3}  \\
& + &  
\frac{2 (1-\theta _{(0)}^2 ) \theta _{(0)}^6+\left(4 \theta _{(0)}^6+3 \theta _{(0)}^4+4 \theta _{(0)}^2+1\right) \theta _{2(0)}^2}{2 \theta_{(0)} \left(1-\theta _{(0)}^2\right) \left(\theta _{(0)}^2+1\right)^3} a_m^2 \left(1-\mu _\mathrm{o}^2\right) . \nonumber
\ea
The degeneracy in the image positions between the absolute value of the spin and its inclination breaks down with the second order corrections if we consider the angular distances measured along the coordinate axes in the POS, since terms proportional to  $a \mu_\mathrm{o}$  appear together with those proportional to $a \sqrt{1-\mu_\mathrm{o}^2}$. On the other hand, when we consider the angular distance from the center, the spin appears only in the form $a \sqrt{1-\mu_\mathrm{o}^2}$.

An image position in $\t = \t_\mathrm{E}  \left\{ \t_{(0)}^\pm + \t_{(1)}^\mathrm{Sch} \ve +\t_{(2)}^\mathrm{Sch} \ve^2 +{\cal{O}}(\ve^3) \right\}$ solves the general form of the lens equation for a spherically symmetric deflector \cite{bo+se06}
\begin{equation}
\label{eqlen1}
r_\mathrm{os} \sin B = r_\mathrm{o} \sin \theta
\cos(\hat\alpha_\mathrm{Sch}
-\theta)-\sqrt{r_\mathrm{s}^2-r_\mathrm{o}^2\sin^2\theta}
\sin(\hat\alpha_\mathrm{Sch} -\theta).
\end{equation}
with $\hat{\alpha}_\mathrm{Sch}$ being the deflection angle for the Schwarzschild black hole and $r_\mathrm{os}=r_\mathrm{o} \cos B \sqrt{r_\mathrm{s}^2 -r_\mathrm{o}^2 \sin B^2}$ the linear path from the source to the observer. The left hand side can be rewritten in terms of $\tan B$ as $ \left\{ \sqrt{r_\mathrm{s}^2-\sin ^2(\theta ) r_\mathrm{o}^2} \cos \left(\theta -\hat{\alpha}_{\text{Sch}}\right)+\left[ 1-\sin \theta  \sin \left(\theta -\hat{\alpha}_{\text{Sch}}\right)\right] r_\mathrm{o}  \right\} \tan B $. The angles describing image positions and deflection in Eq.~(\ref{eqlen1}) are assumed to be positive. The source position $\beta$ should be taken to be positive when studying an image on the same side of the black hole as the source, and negative when studying an image on the opposite side.

\section{Magnification}
\label{sec:magn}

The ratio between the angular area of the image in the observer sky, $d \t_1 d \t_2$, and the angular area of the source in absence of lensing, $d B_1 d B_2$, gives the (signed) luminous amplification of the image, $A$. It can be calculated as the inverse of the Jacobian determinant of the lensing mapping, $J$, 
\ba
A & = & J^{-1} \label{magn1}\\
 & = & \left[ \frac{\partial B_1 \partial B_2}{\partial \t_1 \partial \t_2} \right]^{-1}  \label{magn2} .
\ea
For a source emitting isotropically, the unlensed source as seen by the observer is $(r_\mathrm{s}/r_\mathrm{os})^2$ smaller than as seen by an observer in the black hole position \cite{boz+al05}. Then,
\beq
J = \left( \frac{r_\mathrm{s}}{r_\mathrm{os}} \right)^{2} \left[ \frac{\partial \mu_\mathrm{s} \partial \phi_\mathrm{s}}{\partial \t_1 \partial \t_2} \right] .  \label{magn3}
\eeq
The Jacobian can be written as a Taylor expansion in $\ve$. We first write the angular position of the source $\{ \mu_\mathrm{s}, \phi_\mathrm{s}\}$  in terms of the angular variable in the POS and then derive with respect to $\t_1$ and $\t_2$. Finally, we introduce the scaled angular variables and rearrange the result as a series expansion in $\ve$. We get
\ba
J   & = & 1 - \frac{1}{ \theta _{(0)}^4}  +  \left\{ \frac{15 (1-\theta _{(0)}^2)^2 \pi }{16 \theta _{(0)}^5 (1+ \theta _{(0)}^2 )} -\frac{4 \theta _{1(0)}
 a_m\sqrt{1-\mu _\mathrm{o}^2}}{\theta _{(0)}^4 (1+ \theta _{(0)}^2 )} \right\} \ve  \label{magn4} \\
& - & 
\left\{ \frac{8 (1-\theta _{(0)}^2 )}{(r_\mathrm{o}+r_\mathrm{s})^2 \theta _{(0)}^6 (1+\theta _{(0)}^2)}  \left[  ( \theta _{(0)}^8+2 \theta_{(0)}^6+2 \theta _{(0)}^4+1 ) r_\mathrm{o}^2
-r_\mathrm{s}^2 ( \theta _{(0)}^8+2 \theta_{(0)}^6+4 \theta _{(0)}^4-8 \theta _{(0)}^2+3\right. \right. \nonumber \\
& -& \left.  2 r_\mathrm{s} r_\mathrm{o}\theta _{(0)}^2 (\theta _{(0)}^4+2 \theta _{(0)}^2-4 ) \right] + 
\frac{225 \pi ^2}{256} \frac{ ( 1-\theta _{(0)}^2 ) (1 -5 \theta _{(0)}^4-2 \theta _{(0)}^2 )}{\theta _{(0)}^6 (1+\theta_{(0)}^2)^3} \nonumber \\
&- & \frac{5 a_m \left(-12 \theta _{(0)}^8+27 \theta _{(0)}^6+7 \theta _{(0)}^4-7 \theta _{(0)}^2+1\right) \theta _{1(0)} \sqrt{1-\mu _\mathrm{o}^2} \pi }{16 \theta _{(0)}^7 \left(\theta _{(0)}^2+1\right)^3} \nonumber \\
& - & \left.   \frac{2 a_m^2 \left(\theta _{(0)}^2 \left(\theta _{(0)}^2+1\right)^2+2 \left(1-3 \theta _{(0)}^4\right) \theta _{1(0)}^2\right) \left(1-\mu _\mathrm{o}^2\right)}{\theta
   _{(0)}^4 \left(1-\theta _{(0)}^2\right) \left(\theta _{(0)}^2+1\right)^3}
\right\} \ve ^2  +{\cal{O}}(\ve^3).  \nonumber
\ea
The corresponding magnification is
\ba
A & = & \frac{\theta _{(0)}^4}{\theta _{(0)}^4-1} 
- \left\{\frac{15 \theta _{(0)}^3 \pi }{16 (1+ \theta _{(0)}^2 )^3} +\frac{4 a_m \theta _{(0)}^4 \theta _{1(0)} \sqrt{1-\mu _\mathrm{o}^2}}{( 1 - \theta
   _{(0)}^2 )^2 (1+\theta _{(0)}^2 )^3}\right\} \ve
\label{magn5}
\\
& -& \left\{ \frac{\theta _{(0)}^2}{ (1-\theta _{(0)}^2 ) (1+ \theta _{(0)}^2)^5} \left[ \frac{675 \theta _{(0)}^4 \pi ^2}{128} -\frac{8 (1+\theta_{(0)}^2)^2}{(r_\mathrm{o}+r_\mathrm{s})^2}  ( \theta _{(0)}^8+2 \theta _{(0)}^6+2 \theta _{(0)}^4+1 ) r_\mathrm{o}^2 \right. \right. \nonumber \\
& -& \left. 2 r_\mathrm{s} r_\mathrm{o} \theta _{(0)}^2 (\theta _{(0)}^4+2 \theta_{(0)}^2-4 ) 
- r_\mathrm{s}^2 (\theta _{(0)}^8+2 \theta _{(0)}^6+4 \theta _{(0)}^4-8 \theta _{(0)}^2+3) \right] \nonumber \\
& + & 
\frac{5 \theta _{(0)} (1 -12 \theta _{(0)}^8+27 \theta _{(0)}^6-17 \theta _{(0)}^4+17 \theta _{(0)}^2 ) \theta _{1(0)} \pi }{16 (1-\theta_{(0)}^2 )^2 (\theta _{(0)}^2+1 )^5}  a_m \sqrt{1-\mu _\mathrm{o}^2}  \nonumber \\
& +& \left. \left [ \theta _{(0)}^2-\frac{6 (1+ \theta _{(0)}^4 ) \theta _{1(0)}^2}{ (1+ \theta_{(0)}^2 )^2} \right]  \frac{2  \theta _{(0)}^4 }{ (1-\theta _{(0)}^2 )^3 (\theta _{(0)}^2+1 )^3} a_m^2  (1-\mu _\mathrm{o}^2)    \right\}\ve^2  +{\cal{O}}(\ve^3).  \nonumber
\ea
The luminous amplification depends on the angular momentum only through terms proportional to $a \sqrt{1-\mu_\mathrm{o}^2}$.

\section{Critical curves and caustics}
\label{sec:crit}

Critical curves are the locus of all images with formally infinite magnification. Points in the lens plane are critical when the Jacobian is singular, $J =0$. We look for a parametric solution in the form
\ba
\t_1 & = & \t_\mathrm{E} \cos \varphi \left\{ 1 + \delta \t_{E,(1)} (\varphi ) \ve + \delta \t_{E,(2)} (\varphi ) \ve^2  \right\},\label{crit1}  \\
\t_2 & = & \t_\mathrm{E}  \sin \varphi \left\{ 1 + \delta \t_{E,(1)} (\varphi ) \ve + \delta \t_{E,(2)} (\varphi ) \ve^2  \right\} ,\label{crit2}
\ea
where $\varphi$ is the polar angle in the POS, i.e. $\tan \varphi = \tan \t_2/\tan \t_1$. In Eqs.~(\ref{crit1},~\ref{crit2}), we have already considered that the critical curve for the Schwarzschild black hole is a circle of radius equal to the Einstein radius $\t_\mathrm{E}$. The condition $J=0$ is fulfilled order by order when
\ba
 \delta \t_{E,(1)} (\varphi ) & = & \frac{15 \pi }{32} +  a_m \sqrt{1-\mu _\mathrm{o}^2} \cos \varphi   , \label{crit3} \\
\delta \t_{E,(2)} (\varphi ) & = &  -\frac{675 \pi ^2}{2048} + 4(1+ D) -\frac{4}{3} D^2 \cos (4 \varphi ) + \frac{15}{32} \pi   a_m \sqrt{1-\mu _\mathrm{o}^2}\cos \varphi  -\frac{1}{2}  a_m^2 \left(1-\mu _\mathrm{o}^2\right)\sin ^2\varphi    .\label{crit4} 
\ea
The critical curve corresponding to the above equations is a circle in the plane $\left\{ \tan \t_1, \tan \t_2 \right\}$. With respect the static black hole, the circle centre is displaced along the equatorial direction by 
\beq
\label{crit5}
 \frac{m a \sqrt{1-\mu _\mathrm{o}^2}}{r_\mathrm{o}} \left( 1+\frac{15 \pi}{32}\ve +{\cal{O}}(\ve^2)      \right) ;
\eeq
the angular momentum does not contribute to the radius, which can be written as
\beq
\label{crit6}
\t_\mathrm{E} \left\{ 1 + \frac{15\pi}{32} \varepsilon +\left[ 4 \left( 1+ D+ D^2 \right)
-\frac{675 \pi ^2}{2048}    \right] \varepsilon ^2   + {\cal{O}}(\ve^3) \right\}
\eeq
Note that $4 D \ve^2 = m/r_\mathrm{o}$.

The corresponding locations in the source plane are the caustics. Given the circular symmetry of the critical curve, the caustics will be point-like and centered in 
\beq
\left\{ B_1 , B_2 \right\} \simeq \left\{ 4 D  a_m \sqrt{1-\mu _\mathrm{o}^2} \varepsilon ^2 \left( 1+ \frac{5 \pi }{16} \ve   + {\cal{O}}(\ve^2)   \right), 0 \right\} .
\eeq
At this leading order in $B$, the tangent can be approximated by the angle. 

At the first order correction in $a$, a circle whose radius is equal to the critical radius in the spherically symmetric case and displaced from the black hole along the equatorial direction by a distance $(\delta-a\sqrt{1-\mu_\mathrm{o}}) \sim \ve^2$ maps onto a circle in the source plane displaced by the same amount $(\delta-a\sqrt{1-\mu_\mathrm{o}})$ and of radius $\delta$ \cite{ra+bl94}. At the next order in $a$, circles map in circles only for displacements of higher order, $(\delta- a\sqrt{1-\mu_\mathrm{o}}) \sim \ve^3$.

\section{Conclusions}
\label{sec:conc}

In this paper we have addressed the study of gravitational lensing by a Kerr black hole in the weak deflection limit. Lensing by rotating objects has been considered a number of times in the past and with very different approaches. Here, we built up the lens equations starting from the geodesics for light rays and then solved for the lensing quantities with a standard perturbative technique. This method allowed us to consider corrections proportional to $a^2 r_\mathrm{Sch}$ and $a r_\mathrm{Sch}^2$. We showed as pure spin terms $\propto a^2,~a^3$ do not contribute to the observable lensing quantities, in particular to the deflection angle. 
 
Up to the first order correction in the spin, the Kerr lens is equivalent to a displaced Schwarzschild deflector. This is a very general property of spinning lenses \cite{ser02}. To the next order, this degeneracy is broken and some particular features show up. The two perturbed images are no more aligned with a fixed position. The degeneracy between the absolute value of the spin and its inclination on the line of sight is also broken. All observable quantities at the first order correction in the spin are functions of $a \sin \vt_\mathrm{o}$ but terms proportional to $a\cos \vt_\mathrm{o}$ appear at the next order in the angular coordinates of the images in the plane of the sky. However, the angular displacement of the images from the center is still a function of $a \sin \vt_\mathrm{o}$.
 The shape of the critical curve is still a circle displaced along the equatorial direction and the caustic is still point-like. The finite size of the caustic should show up at the next order due to terms $\propto a^2 r_\mathrm{Sch}^2$ as suggested by numerical results \cite{ra+bl94}.

It could be of interest to draw some comparison with the case of the strong deflection limit \cite{boz03,boz+al05,boz+al06}. Such a limit has been treated considering small values of the angular momentum and including corrections proportional to $a^2$. That is two orders beyond the Schwarzschild lens. This was enough to obtain finite shaped caustics. In the present study of the weak deflection limit, we made no assumptions on the absolute value of the spin and still considered two orders beyond the spherically symmetric lens but we did non get the caustic structure. This is only an apparent discrepancy, as we have to remind that the minimum distance in the strong deflection limit is of order of the gravitational radius. In fact the finite size of the caustic springs from terms proportional to $(a^2 r_\mathrm{Sch}^2)/r_\mathrm{min}^4$. Since in the strong deflection limit $r_\mathrm{min} \sim r_\mathrm{Sch}$, we see as these terms are included in an analysis at the second order in $a$.

If the supermassive black hole at the Galactic center has a significant angular momentum, some features of Kerr lensing could be detected by future space astrometric mission with a planned resolution of the microarcsecond. 

In a forthcoming paper we will present an analytical treatment of Kerr lensing in the weak deflection limit accounting for the caustic structure.

\begin{acknowledgments}
M.S. is supported by the Swiss National Science Foundation and by the Tomalla Foundation.  F.D.L.'s work was performed under the auspices of the EU, which has provided financial support to the `Dottorato di Ricerca Internazionale in Fisica della Gravitazione ed Astrofisica' of the Salerno University, through `Fondo Sociale Europeo, Misura III.4'.
\end{acknowledgments}

\appendix

\section{Radial integrals}
\label{app:a}

These appendices are devoted to the resolution of the integrals in the geodesic equations. Let us start with some considerations on the radial integrals. The sign convention in the geodesic equations remind us that integrations must be performed summing up with the same sign all contributions from paths bounded by consecutive inversion points. For a standard gravitational lensing configuration, the $r$-motion, $r_\mathrm{s} \rightarrow r_\mathrm{min} \rightarrow r_\mathrm{o}$, has only one inversion point so that we have to add the contributions due to the approach and the departure of the photon. Integrals can be easily evaluated expanding the integrand as a Taylor series in $\epsilon$ and then performing the integration term by term. When evaluating the expanded primitive function in the extrema $r_\mathrm{s}$ and $r_\mathrm{o}$, we remind that  $\sqrt{J^2+Q}/r_\mathrm{s}$ and $\sqrt{J^2+Q}/r_\mathrm{o}$ are of order of $\epsilon$. Let us first consider the left-hand side of Eq.~(\ref{ker4}). The integral reads
\begin{eqnarray}
 \int_{r_\mathrm{min}}^{r_\mathrm{s}} \frac{d r}{\sqrt{R}}+\int_{r_\mathrm{min}}^{r_\mathrm{o}} \frac{d r}{\sqrt{R}} & \simeq &
\frac{\pi }{\sqrt{J^2+Q}}+\frac{2}{J^2+Q}r_\text{Sch} \\
& +& \frac{15 \pi  r_{\text{Sch}}^2}{16 \left(J^2+Q\right)^{3/2}}-\frac{4 J a r_{\text{Sch}}}{\left(J^2+Q\right)^2}
+\frac{a^2 \pi  \left(2 J^2-Q\right)}{4 \left(J^2+Q\right)^{5/2}} \nonumber \\
& + & \frac{16 r_{\text{Sch}}^3}{3 \left(J^2+Q\right)^2}-\frac{15 a J \pi  r_{\text{Sch}}^2}{4 \left(J^2+Q\right)^{5/2}}+\frac{a^4 \left(6 J^2-2 Q\right) r_{\text{Sch}}}{\left(J^2+Q\right)^3} \nonumber \\
& - & \frac{J^2+Q}{6 r_\mathrm{o}^3}-\frac{J^2+Q}{6 r_\mathrm{s}^3}-\frac{1}{r_0}-\frac{1}{r_\mathrm{s}}    \nonumber .
\end{eqnarray}

Let us now consider the radial integral in the right hand side of Eq.~(\ref{ker5}). We have
\begin{eqnarray}
 \int_{r_\mathrm{min}}^{r_\mathrm{s}} \frac{a \left( r_\text{Sch}r -a J\right )}{\pm \Delta{\sqrt{R}} }dr
+\int_{r_\mathrm{min}}^{r_\mathrm{o}} \frac{a \left( r_\text{Sch} r -a J\right )}{\pm \Delta{\sqrt{R}} }dr
& \simeq & \frac{2 r_{\text{Sch}} a}{J^2+Q} -\frac{\pi  a^2}{2 \left(J^2+Q\right)^{3/2}} \label{intr2} \\
& - & \frac{4 J r_{\text{Sch}} a^2}{\left(J^2+Q\right)^2} 
 +  \frac{5 \pi  r_{\text{Sch}}^2 a}{4 \left(J^2+Q\right)^{3/2}} . \nonumber
\end{eqnarray}
Corrections due to the finiteness of $r_\mathrm{o}$ and $r_\mathrm{s}$ in Eq.~(\ref{intr2}) of order $\sim \epsilon^4$.  We remark as radial integrals can be more easily solved changing to the variable $u = r_\mathrm{min}/r $.

\section{Angular integrals}
\label{app:b}

The angular integrals follow the photon polar trajectory from the source, $\vartheta_\mathrm{s}$, to the turning point, which is either the minimum $\vartheta_\mathrm{min}$ or the maximum $\vartheta_\mathrm{max}$, to the observer at $\vartheta_\mathrm{o}$. As for the radial motion, the integration is a path integral over the whole trajectory of the photon with all contributions to be summed with the same sign. The two branches, i.e $\vt_\mathrm{s} \rightarrow ( \vt_\mathrm{min}, \vt_\mathrm{max}) \rightarrow \vt_\mathrm{o} $ sum up positively if we take the sign of $\sqrt{\Theta}$ to be positive (negative) if integrating from $\vartheta_\mathrm{s}$ to $\vartheta_\mathrm{max}$ ($\vartheta_\mathrm{min}$) and negative (positive) from $\vartheta_\mathrm{max}$ ($\vartheta_\mathrm{min}$) to  $\vartheta_\mathrm{o}$. It is useful to change to $\mu \equiv \cos \vartheta$. The right hand side of Eq.~(\ref{ker4}) can be then rewritten as
\beq
\label{ang1}
I_\mu=\int \frac{1}{\pm  \sqrt{\Theta_\mu}} d \mu ,  \\
\eeq
where
\begin{eqnarray}
\Theta_\mu &\equiv & a^2(\mu_-^2+\mu^2)(\mu_+^2-\mu^2)  , \label{ang2}  \\
\mu_\pm^2 & \equiv &\frac{\sqrt{b_{JQ}^2+4a^2Q_m}\pm b_{JQ}}{2 a^2}  \label{ang3}  , \\
b_{JQ} &=& a^2-J^2-Q .  \label{ang4}
\end{eqnarray}
The turning point in the polar motion is a zero of $\Theta_\mu$, i.e. $\pm \mu_+$, with $ \mu_+$ corresponding to $\vartheta_\mathrm{min}$. To the lowest orders
\beq
\mu_+ \simeq \sqrt{\frac{Q}{J^2 + Q}}\left[  \frac{a^2 J^2}{2 \left(J^2 +Q\right)^2}+ \frac{a^4 J^2 (3 J^2 - 4 Q)}{8 \left(J^2 + Q\right)^4} + {\cal{O}}(\ve^6)  \right] .
\eeq
The primitive function of the integral in Eq.~(\ref{ang1}) is
\beq
\label{ang5}
PI(\mu) = \frac{1}{a \mu _-}{F \left( \sin ^{-1}\left(\mu/\mu_+ \right),-(\mu _+/\mu_-)^2\right) + {\cal{O}}(\ve^6) } ,
\eeq
where $F$ is the elliptic integral of the first kind. We remark as the integral in Eq.~(\ref{ang1}) can be more easily solved in terms of the integration variable $\mu/\mu_+$ \cite{cha83,bra86,boz+al06}. The function in Eq.~(\ref{ang5}) can be expanded as
\beq
\label{ang6}
PI(\mu) \simeq \frac{1}{\sqrt{J^2+Q}} \left\{ \sin ^{-1}\mu_\sigma - 
\left[  \frac{2 J^2 -Q(1 - \mu_\sigma^2) }{4 (J^2 + Q )} \frac{\mu_\sigma}{\sqrt{1 -\mu_\sigma^2}} 
+\frac{Q - 2 J^2}{4(J^2 + Q)}\sin^{-1}\mu_\sigma \right] \frac{a^2}{J^2 + Q}\right\} ,
\eeq
with $\mu_\sigma \equiv \mu /\sqrt{Q/(J^2+Q)}$. In $\mu =\mu_+$, the primitive function reduces to
\beq
\label{ang7}
PI(\mu_+) \simeq \frac{\pi }{2 \sqrt{J^2 + Q}}+ \frac{a^2 \pi  \left(2 J^2 - Q\right)}{8 \left(J^2 + Q\right)^{5/2}}.
\eeq
The turning point is attained in either $\mu_+$ or $-\mu_+$, according to that photon gets a minimum or a maximum polar angle, respectively. We remind that $\mu$ is a decreasing function of $\vartheta$ so that the considerations on the signs must be accordingly updated. Using the property that  $PI(-\mu_+) = -PI(\mu_+) $ and following the sign convention sum, we sum up the paths as
\beq
\label{ang8}
I_\mu = 2 PI(\mu_+) +(-1)^k [PI(\mu_\mathrm{s})+ PI(\mu_\mathrm{o}) ],
\eeq
with $k$ an integer number defined to be even (odd) if the photon gets to the observer from below (from the top), i.e. after having reached $\vartheta_\mathrm{max}$ ($\vartheta_\mathrm{min}$).

Let us now consider the angular integral in the right hand side of Eq.~(\ref{ker5}),
\beq
\label{ang10}
J_\vt= \int \frac{csc^2\vartheta}{\pm \sqrt{\Theta}} d \vartheta.
\eeq
In terms of $\mu$, the above integral can be written as
\beq
\label{ang11}
J_\mu = \int \frac{1}{\pm (1-\mu^2)\sqrt{\Theta_\mu}} d \mu;
\eeq
the primitive function can be expressed in terms of the incomplete elliptic integral of the third kind, 
\beq
\label{ang12}
PJ(\mu) = -\frac{J \Pi \left(\mu _+^2;-\sin ^{-1}\left(\frac{\mu }{\mu _+}\right)|-\frac{\mu _+^2}{\mu _-^2}\right)}{a \mu _-} .
\eeq
To the lowest orders, Eq.~(\ref{ang12}) reduces to
\beq
\label{ang13}
PJ(\mu) \simeq  \tan ^{-1}\left(\frac{J \mu_\sigma }{\sqrt{J^2+Q} \sqrt{1-\mu_\sigma^2}}\right)-\frac{a^2 J }{2 \left(J^2+Q\right)^{3/2}}\left(\frac{\mu_\sigma}{\sqrt{1-\mu_\sigma^2}}-\sin ^{-1}\mu_\sigma  \right)  .
\eeq
In  $\mu =\mu_+$,
\beq
\label{ang14}
PJ(\mu_+) \simeq \frac{J \pi  a^2}{4 \left(J^2+Q\right)^{3/2}}+\frac{J \pi }{2 |J|} .
\eeq
The sum convention works as in Eq.~(\ref{ang8}).

%\bibliography{kerr_weak}

\begin{thebibliography}{32}
\expandafter\ifx\csname natexlab\endcsname\relax\def\natexlab#1{#1}\fi
\expandafter\ifx\csname bibnamefont\endcsname\relax
  \def\bibnamefont#1{#1}\fi
\expandafter\ifx\csname bibfnamefont\endcsname\relax
  \def\bibfnamefont#1{#1}\fi
\expandafter\ifx\csname citenamefont\endcsname\relax
  \def\citenamefont#1{#1}\fi
\expandafter\ifx\csname url\endcsname\relax
  \def\url#1{\texttt{#1}}\fi
\expandafter\ifx\csname urlprefix\endcsname\relax\def\urlprefix{URL }\fi
\providecommand{\bibinfo}[2]{#2}
\providecommand{\eprint}[2][]{\url{#2}}

\bibitem[{\citenamefont{{Bozza}}(2002)}]{boz02}
\bibinfo{author}{\bibfnamefont{V.}~\bibnamefont{{Bozza}}},
  \bibinfo{journal}{\prd} \textbf{\bibinfo{volume}{66}},
  \bibinfo{pages}{103001} (\bibinfo{year}{2002}), \eprint{gr-qc/0208075}.

\bibitem[{\citenamefont{{Keeton} and {Petters}}(2005)}]{ke+pe05}
\bibinfo{author}{\bibfnamefont{C.~R.} \bibnamefont{{Keeton}}} \bibnamefont{and}
  \bibinfo{author}{\bibfnamefont{A.~O.} \bibnamefont{{Petters}}},
  \bibinfo{journal}{\prd} \textbf{\bibinfo{volume}{72}},
  \bibinfo{pages}{104006} (\bibinfo{year}{2005}).

\bibitem[{\citenamefont{{Amore} and {Arceo}}(2006)}]{am+ar06}
\bibinfo{author}{\bibfnamefont{P.}~\bibnamefont{{Amore}}} \bibnamefont{and}
  \bibinfo{author}{\bibfnamefont{S.}~\bibnamefont{{Arceo}}},
  \bibinfo{journal}{\prd} \textbf{\bibinfo{volume}{73}},
  \bibinfo{pages}{083004} (\bibinfo{year}{2006}), \eprint{gr-qc/0602106}.

\bibitem[{\citenamefont{{Carter}}(1968)}]{car68}
\bibinfo{author}{\bibfnamefont{B.}~\bibnamefont{{Carter}}},
  \bibinfo{journal}{Physical Review} \textbf{\bibinfo{volume}{174}},
  \bibinfo{pages}{1559} (\bibinfo{year}{1968}).

\bibitem[{\citenamefont{{Chandrasekhar}}(1983)}]{cha83}
\bibinfo{author}{\bibfnamefont{S.}~\bibnamefont{{Chandrasekhar}}},
  \emph{\bibinfo{title}{{The mathematical theory of black holes}}}
  (\bibinfo{publisher}{Clarendon, Oxford}, \bibinfo{year}{1983}).

\bibitem[{\citenamefont{{Cunningham} and {Bardeen}}(1973)}]{cu+ba73}
\bibinfo{author}{\bibfnamefont{J.~M.} \bibnamefont{{Cunningham}}}
  \bibnamefont{and} \bibinfo{author}{\bibfnamefont{C.~T.}
  \bibnamefont{{Bardeen}}}, \bibinfo{journal}{\apj}
  \textbf{\bibinfo{volume}{183}}, \bibinfo{pages}{237} (\bibinfo{year}{1973}).

\bibitem[{\citenamefont{{Viergutz}}(1993)}]{vie93}
\bibinfo{author}{\bibfnamefont{S.~U.} \bibnamefont{{Viergutz}}},
  \bibinfo{journal}{Astron. Astroph.} \textbf{\bibinfo{volume}{272}},
  \bibinfo{pages}{355} (\bibinfo{year}{1993}).

\bibitem[{\citenamefont{{Rauch} and {Blandford}}(1994)}]{ra+bl94}
\bibinfo{author}{\bibfnamefont{K.~P.} \bibnamefont{{Rauch}}} \bibnamefont{and}
  \bibinfo{author}{\bibfnamefont{R.~D.} \bibnamefont{{Blandford}}},
  \bibinfo{journal}{\apj} \textbf{\bibinfo{volume}{421}}, \bibinfo{pages}{46}
  (\bibinfo{year}{1994}).

\bibitem[{\citenamefont{{Bozza}}(2003)}]{boz03}
\bibinfo{author}{\bibfnamefont{V.}~\bibnamefont{{Bozza}}},
  \bibinfo{journal}{\prd} \textbf{\bibinfo{volume}{67}},
  \bibinfo{pages}{103006} (\bibinfo{year}{2003}), \eprint{gr-qc/0210109}.

\bibitem[{\citenamefont{{Bozza} et~al.}(2005)\citenamefont{{Bozza}, {De Luca},
  {Scarpetta}, and {Sereno}}}]{boz+al05}
\bibinfo{author}{\bibfnamefont{V.}~\bibnamefont{{Bozza}}},
  \bibinfo{author}{\bibfnamefont{F.}~\bibnamefont{{De Luca}}},
  \bibinfo{author}{\bibfnamefont{G.}~\bibnamefont{{Scarpetta}}},
  \bibnamefont{and} \bibinfo{author}{\bibfnamefont{M.}~\bibnamefont{{Sereno}}},
  \bibinfo{journal}{\prd} \textbf{\bibinfo{volume}{72}},
  \bibinfo{pages}{083003} (\bibinfo{year}{2005}), \eprint{gr-qc/0507137}.

\bibitem[{\citenamefont{{Bozza} et~al.}(2006)\citenamefont{{Bozza}, {De Luca},
  and {Scarpetta}}}]{boz+al06}
\bibinfo{author}{\bibfnamefont{V.}~\bibnamefont{{Bozza}}},
  \bibinfo{author}{\bibfnamefont{F.}~\bibnamefont{{De Luca}}},
  \bibnamefont{and}
  \bibinfo{author}{\bibfnamefont{G.}~\bibnamefont{{Scarpetta}}},
  \bibinfo{journal}{ArXiv:gr-qc/0604093}  (\bibinfo{year}{2006}).

\bibitem[{\citenamefont{{Pineault} and {Roeder}}(1977)}]{pi+ro77}
\bibinfo{author}{\bibfnamefont{S.}~\bibnamefont{{Pineault}}} \bibnamefont{and}
  \bibinfo{author}{\bibfnamefont{R.~C.} \bibnamefont{{Roeder}}},
  \bibinfo{journal}{\apj} \textbf{\bibinfo{volume}{212}}, \bibinfo{pages}{541}
  (\bibinfo{year}{1977}).

\bibitem[{\citenamefont{Epstein and Shapiro}(1980)}]{ep+sh80}
\bibinfo{author}{\bibfnamefont{R.}~\bibnamefont{Epstein}} \bibnamefont{and}
  \bibinfo{author}{\bibfnamefont{I.~I.} \bibnamefont{Shapiro}},
  \bibinfo{journal}{Phys. Rev. D} \textbf{\bibinfo{volume}{22}},
  \bibinfo{pages}{2947} (\bibinfo{year}{1980}).

\bibitem[{\citenamefont{Ib\'a\~nez and Mart\'in}(1982)}]{ib+ma82}
\bibinfo{author}{\bibfnamefont{J.}~\bibnamefont{Ib\'a\~nez}} \bibnamefont{and}
  \bibinfo{author}{\bibfnamefont{J.}~\bibnamefont{Mart\'in}},
  \bibinfo{journal}{Phys. Rev. D} \textbf{\bibinfo{volume}{26}},
  \bibinfo{pages}{384} (\bibinfo{year}{1982}).

\bibitem[{\citenamefont{{Ib\'a\~nez}}(1983)}]{iba83}
\bibinfo{author}{\bibfnamefont{J.}~\bibnamefont{{Ib\'a\~nez}}},
  \bibinfo{journal}{Astron. Astroph.} \textbf{\bibinfo{volume}{124}},
  \bibinfo{pages}{175} (\bibinfo{year}{1983}).

\bibitem[{\citenamefont{{Bray}}(1986)}]{bra86}
\bibinfo{author}{\bibfnamefont{I.}~\bibnamefont{{Bray}}},
  \bibinfo{journal}{\prd} \textbf{\bibinfo{volume}{34}}, \bibinfo{pages}{367}
  (\bibinfo{year}{1986}).

\bibitem[{\citenamefont{{Dymnikova}}(1986)}]{dym86}
\bibinfo{author}{\bibfnamefont{I.~G.} \bibnamefont{{Dymnikova}}}, in
  \emph{\bibinfo{booktitle}{IAU Symp. 114: Relativity in Celestial Mechanics
  and Astrometry. High Precision Dynamical Theories and Observational
  Verifications}}, edited by
  \bibinfo{editor}{\bibfnamefont{J.}~\bibnamefont{{Kovalevsky}}}
  \bibnamefont{and} \bibinfo{editor}{\bibfnamefont{V.~A.}
  \bibnamefont{{Brumberg}}} (\bibinfo{year}{1986}), p. \bibinfo{pages}{411}.

\bibitem[{\citenamefont{{Glicenstein}}(1999)}]{gli99}
\bibinfo{author}{\bibfnamefont{J.~F.} \bibnamefont{{Glicenstein}}},
  \bibinfo{journal}{Astron. Astroph.} \textbf{\bibinfo{volume}{343}},
  \bibinfo{pages}{1025} (\bibinfo{year}{1999}).

\bibitem[{\citenamefont{{Kopeikin}}(1997)}]{kop97}
\bibinfo{author}{\bibfnamefont{S.}~\bibnamefont{{Kopeikin}}},
  \bibinfo{journal}{J. Math. Phys.} \textbf{\bibinfo{volume}{38}},
  \bibinfo{pages}{2587} (\bibinfo{year}{1997}).

\bibitem[{\citenamefont{{Kopeikin} and {Sch{\"a}fer}}(1999)}]{ko+sc99}
\bibinfo{author}{\bibfnamefont{S.}~\bibnamefont{{Kopeikin}}} \bibnamefont{and}
  \bibinfo{author}{\bibfnamefont{G.}~\bibnamefont{{Sch{\"a}fer}}},
  \bibinfo{journal}{\prd} \textbf{\bibinfo{volume}{60}},
  \bibinfo{pages}{124002} (\bibinfo{year}{1999}), \eprint{gr-qc/9902030}.

\bibitem[{\citenamefont{{Kopeikin} and {Mashhoon}}(2002)}]{ko+ma02}
\bibinfo{author}{\bibfnamefont{S.}~\bibnamefont{{Kopeikin}}} \bibnamefont{and}
  \bibinfo{author}{\bibfnamefont{B.}~\bibnamefont{{Mashhoon}}},
  \bibinfo{journal}{\prd} \textbf{\bibinfo{volume}{65}},
  \bibinfo{pages}{064025} (\bibinfo{year}{2002}), \eprint{gr-qc/0110101}.

\bibitem[{\citenamefont{{Asada} and {Kasai}}(2000)}]{as+ka00}
\bibinfo{author}{\bibfnamefont{H.}~\bibnamefont{{Asada}}} \bibnamefont{and}
  \bibinfo{author}{\bibfnamefont{M.}~\bibnamefont{{Kasai}}},
  \bibinfo{journal}{Progress of Theoretical Physics}
  \textbf{\bibinfo{volume}{104}}, \bibinfo{pages}{95} (\bibinfo{year}{2000}),
  \eprint{astro-ph/0006157}.

\bibitem[{\citenamefont{{Sereno}}(2002)}]{ser02}
\bibinfo{author}{\bibfnamefont{M.}~\bibnamefont{{Sereno}}},
  \bibinfo{journal}{Physics Letters A} \textbf{\bibinfo{volume}{305}},
  \bibinfo{pages}{7} (\bibinfo{year}{2002}), \eprint{astro-ph/0209148}.

\bibitem[{\citenamefont{{Sereno} and {Cardone}}(2002)}]{se+ca02}
\bibinfo{author}{\bibfnamefont{M.}~\bibnamefont{{Sereno}}} \bibnamefont{and}
  \bibinfo{author}{\bibfnamefont{V.~F.} \bibnamefont{{Cardone}}},
  \bibinfo{journal}{Astron. Astroph.} \textbf{\bibinfo{volume}{396}},
  \bibinfo{pages}{393} (\bibinfo{year}{2002}), \eprint{astro-ph/0209297}.

\bibitem[{\citenamefont{{Sereno}}(2003{\natexlab{a}})}]{ser03}
\bibinfo{author}{\bibfnamefont{M.}~\bibnamefont{{Sereno}}},
  \bibinfo{journal}{MNRAS} \textbf{\bibinfo{volume}{344}}, \bibinfo{pages}{942}
  (\bibinfo{year}{2003}{\natexlab{a}}), \eprint{astro-ph/0307243}.

\bibitem[{\citenamefont{{Sereno}}(2003{\natexlab{b}})}]{ser03b}
\bibinfo{author}{\bibfnamefont{M.}~\bibnamefont{{Sereno}}},
  \bibinfo{journal}{\prd} \textbf{\bibinfo{volume}{67}},
  \bibinfo{pages}{064007} (\bibinfo{year}{2003}{\natexlab{b}}),
  \eprint{astro-ph/0301290}.

\bibitem[{\citenamefont{{Capozziello} et~al.}(2003)\citenamefont{{Capozziello},
  {Cardone}, {Re}, and {Sereno}}}]{cap+al03}
\bibinfo{author}{\bibfnamefont{S.}~\bibnamefont{{Capozziello}}},
  \bibinfo{author}{\bibfnamefont{V.~F.} \bibnamefont{{Cardone}}},
  \bibinfo{author}{\bibfnamefont{V.}~\bibnamefont{{Re}}}, \bibnamefont{and}
  \bibinfo{author}{\bibfnamefont{M.}~\bibnamefont{{Sereno}}},
  \bibinfo{journal}{MNRAS} \textbf{\bibinfo{volume}{343}}, \bibinfo{pages}{360}
  (\bibinfo{year}{2003}), \eprint{astro-ph/0304272}.

\bibitem[{\citenamefont{{Sereno}}(2005)}]{ser05}
\bibinfo{author}{\bibfnamefont{M.}~\bibnamefont{{Sereno}}},
  \bibinfo{journal}{MNRAS} \textbf{\bibinfo{volume}{357}},
  \bibinfo{pages}{1205} (\bibinfo{year}{2005}), \eprint{astro-ph/0412108}.

\bibitem[{\citenamefont{{Bodenner} and {Will}}(2003)}]{bo+wi03}
\bibinfo{author}{\bibfnamefont{J.}~\bibnamefont{{Bodenner}}} \bibnamefont{and}
  \bibinfo{author}{\bibfnamefont{C.~M.} \bibnamefont{{Will}}},
  \bibinfo{journal}{Am. J. Phys.} \textbf{\bibinfo{volume}{71}},
  \bibinfo{pages}{770} (\bibinfo{year}{2003}).

\bibitem[{\citenamefont{{Asada} et~al.}(2003)\citenamefont{{Asada}, {Kasai},
  and {Yamamoto}}}]{asa+al03}
\bibinfo{author}{\bibfnamefont{H.}~\bibnamefont{{Asada}}},
  \bibinfo{author}{\bibfnamefont{M.}~\bibnamefont{{Kasai}}}, \bibnamefont{and}
  \bibinfo{author}{\bibfnamefont{T.}~\bibnamefont{{Yamamoto}}},
  \bibinfo{journal}{\prd} \textbf{\bibinfo{volume}{67}},
  \bibinfo{pages}{043006} (\bibinfo{year}{2003}), \eprint{astro-ph/0301099}.

\bibitem[{\citenamefont{{Eisenhauer} et~al.}(2005)\citenamefont{{Eisenhauer},
  {Genzel}, {Alexander}, {Abuter}, {Paumard}, {Ott}, {Gilbert}, {Gillessen},
  {Horrobin}, {Trippe} et~al.}}]{eis+al05}
\bibinfo{author}{\bibfnamefont{F.}~\bibnamefont{{Eisenhauer}}},
  \bibinfo{author}{\bibfnamefont{R.}~\bibnamefont{{Genzel}}},
  \bibinfo{author}{\bibfnamefont{T.}~\bibnamefont{{Alexander}}},
  \bibinfo{author}{\bibfnamefont{R.}~\bibnamefont{{Abuter}}},
  \bibinfo{author}{\bibfnamefont{T.}~\bibnamefont{{Paumard}}},
  \bibinfo{author}{\bibfnamefont{T.}~\bibnamefont{{Ott}}},
  \bibinfo{author}{\bibfnamefont{A.}~\bibnamefont{{Gilbert}}},
  \bibinfo{author}{\bibfnamefont{S.}~\bibnamefont{{Gillessen}}},
  \bibinfo{author}{\bibfnamefont{M.}~\bibnamefont{{Horrobin}}},
  \bibinfo{author}{\bibfnamefont{S.}~\bibnamefont{{Trippe}}},
  \bibnamefont{et~al.}, \bibinfo{journal}{\apj} \textbf{\bibinfo{volume}{628}},
  \bibinfo{pages}{246} (\bibinfo{year}{2005}), \eprint{astro-ph/0502129}.

\bibitem[{\citenamefont{{Bozza} and {Sereno}}(2006)}]{bo+se06}
\bibinfo{author}{\bibfnamefont{V.}~\bibnamefont{{Bozza}}} \bibnamefont{and}
  \bibinfo{author}{\bibfnamefont{M.}~\bibnamefont{{Sereno}}},
  \bibinfo{journal}{\prd} \textbf{\bibinfo{volume}{73}},
  \bibinfo{pages}{103004} (\bibinfo{year}{2006}), \eprint{gr-qc/0603049}.

\end{thebibliography}

\end{document}